# INTERFACING THE CONTROLLOGIX PLC OVER ETHERNET/IP

K.U. Kasemir, L.R. Dalesio, LANL, Los Alamos, NM 87545, USA

Abstract

The Allen-Bradley ControlLogix [1] line of programmable logic controllers (PLCs) offers several interfaces: Ethernet, ControlNet, DeviceNet, RS-232 and others. The ControlLogix Ethernet interface module 1756-ENET uses EtherNet/IP, the ControlNet protocol [2], encapsulated in Ethernet packages, with specific service codes [3]. A driver for the Experimental Physics and Industrial Control System (EPICS) has been developed that utilizes this EtherNet/IP protocol for controllers running the vxWorks RTOS as well as a Win32 and Unix/Linux test program. Features, performance and limitations of this interface are presented.

## 1 INTRODUCTION

Several subsystems of the Spallation Neutron Source project (SNS) employ Allen-Bradley ControlLogix PLCs [4]. To integrate these into the EPICS-based accelerator control system, the EPICS input/output controllers (IOCs) need read and write access to the PLC data. Since the IOCs, their Unix or Win32 boot hosts as well as almost every PC which is used to program the PLC is already equipped with an Ethernet interface, it is desirable to use the same technology for transferring the PLC data. Existing support and knowledge for cabling, network hardware, configuration and maintenance can thus be utilized.

## 2 ETHERNET/IP

ControlNet is a deterministic serial communication system, its specification extends from the physical to the application layer of the seven layer ISO OSI model[5]. ControlNet Release 2.0 [2] introduced the TCP/IP encapsulation of data packages, replacing the Physical and Data Link layer with Ethernet respectively IP/UDP/TCP. The result was known as "ControlNet over Ethernet" and is now called EtherNet/IP[6].

After connecting to an EtherNet/IP target, by default on TCP/IP port 0xAF12, and establishing a session ID via the encapsulation protocol, messages can be exchanged. They are defined in the object oriented Control and Information Protocol (CIP), part of the ControlNet specification. In reference to the ControlNet transport layer, one still distinguishes "unconnected" and "connected" CIP messages. They are encapsulated differently but can both be transmitted via TCP, which – by definition – is always connected.

## 3 CONTROLLOGIX ETHERNET INTERFACE

The ControlLogix PLC uses ControlNet to communicate with local I/O boards over the back plane, the 1756-ENET Ethernet module supports EtherNet/IP. Following the EtherNet/IP specification, one can use the *SendRRData* encapsulation command and send this unconnected CIP message to the interface:
  Service: Get_Attribute_Single (0x0E)
  Path: Identity Object (class 0x01, instance 1),
      Product Name (attrib. 7)
In response, the interface sends a reply:
  Service: Get_Attribute_Single-Reply (0x8E)
  Response: length=12, "1756-ENET/A "

The CIP object model also includes "Analog Input Point" (0x0A) and "Discrete Input Group" (0x1D) objects, but so far our attempts to use these for accessing ControlLogix analog or digital input modules have failed. There is no standard CIP object that suggests usability for accessing tag names on the PLC.

Instead, Allen-Bradley published new CIP service codes specific to ControlLogix [3], including CIP path names for ladder logic tags, service codes for individual read and write access and combined transfers as well as a binary data format used for these transactions.

To send CIP extensions to the PLC with an *SendRRData* encapsulation command, they have to be routed via the *Connection Manager* object of the 1756-ENET module, i.e. embedded like this:
  Service: CM_Unconnected_Send (0x52)
  Path: Connection Manager (class 0x06, instance 1)
  <encoded timeout>, <embedded message>
  Path: Port 1(back plane), Link # (slot # of PLC).
This embedded message reads a tag named 'TEST':
  Service: CIP_Read_Data (0x4C)
  Path: 'TEST' (ANSI extended symbol segment)
  Elements: UINT 1
On success the interface forwards the reply from the PLC, the Connection Manager becomes transparent:
  Service: CIP_Read_Data-Reply (0xCC)
  Response: CA 00 00 80 38 3B = REAL 0.002815
The CIP_Write_Data service (0x53) allows modification of tags on the PLC via similar embedding.

In contrast to other communication protocols, no change to the ladder logic is required! The CIP Read/Write services can access all controller tags with no need to previously mark them as "published" or "consumed" in the PLC programming software. This includes access to I/O modules: The first channel of an analog input module in slot 1 is available as "Local:1:I.Ch0Data".

This type of transfer is called explicit unconnected messaging, because the tag name is explicitly mentioned and each packet is individually routed. For connected messaging, the Message Router on the PLC is instructed to open a connection:

*Service: CM_Forward_Open (0x54*
*Path: Connection Manager (class 0x06, instance 1)*
*<timeout, connection ID, update interval, ...>*
*Connection Path: Port 1, Link 0 (back plane,*
*PLC slot), Message Router (class 0x02, instance 1)*

The reply provides a serial number. The CIP_Read_Data requests can now be sent as connected messages with *SendUnitData* encapsulation, prefixed by a sequence number, without embedding them in a routing CM_Unconnected_Send message.

## 4 DESIGN DECISIONS

Unconnected messaging is used since the advantages of connected messaging do not transfer from ControlNet to EtherNet/IP: Ethernet does not reserve bandwidth; guaranteed delivery is already handled by TCP. For CIP_Read_Data requests, comparison of connected to unconnected messaging resulted in slightly smaller messages and a 3% increase in throughput. As a drawback, the client application has to send requests at the established update interval of the connection or faster. Temporary Ethernet delays cause the PLC to close the connection.

The ControlLogix Multi-Request Service (0x0A) is used to combine CIP_Read/Write_Data requests until either the total request or expected response size reaches the PLC buffer limit of approximately 500 bytes. (Chapter 8.3.1.4 in [2] defines this as 511 bytes, 2-4.1 in [6] as 504 bytes).

## 5 IMPLEMENTATION OF EPICS SUPPORT

For each PLC, the vxWorks driver code arranges the tags in scan lists depending on the requested update rate. One thread per PLC handles all read/write requests.

EPICS device support allows analog, binary and multi-bit records to use the driver for input and output. Tags have to refer to a scalar value, a single array element or a structure element, not whole arrays or structures. The PLC data types BOOL, SINT, INT, DINT and REAL are handled.

One can change the record configuration at runtime, without rebooting the IOC, e.g. the tag name that a record refers to can be replaced. In case of a communication error or timeout, the driver disconnects from the PLC and attempts periodic reconnects.

Per default, the driver combines requests for array elements into one array transfer from the first to the highest requested element. This leads to a significant reduction in transfer times, but might have side effects: The IOC will always write the whole array whenever more than one element has been changed by output records. If the same PLC array has been modified by another source (PanelView display) since the last transfer, the IOC is unaware of these changes and will overwrite them. An array transfer is also size-limited by the aforementioned PLC buffer limit. The record configuration allows separate array element transfers as a workaround for these cases.

For output records, the driver sends a CIP_Write_Data message whenever the record is processed. Otherwise it will periodically read the tag from the PLC and update the output record if the value on the PLC differs from the one in the record.

The driver keeps statistical information (error counts, last/minimum/maximum transfer time) for each scan list. Analog input records allow access to these values.

One problem arose with BOOL arrays since they are transferred as DINT values. For an analog record, a tag of "test[5]" is interpreted as addressing the $5^{th}$ element of tag "test". When this is applied to a BOOL array, the result would be the $5^{th}$ DINT, containing bits 160-191. So for binary records, all array access is assumed to target BOOL arrays, and "test[5]" would be transformed in a request to DINT[0], bit 5.

This software has been tested on 68K, PPC and Pentium IOCs. The lower driver layer handles the different byte order. In addition, a simple command-line program is available for Unix and Win32 that allows read and write access to PLC tags as a debugging aid.

## 6 PERFORMANCE

For the following, an MVME2100 CPU with a 100baseT-network interface communicated to the PLC with its 10baseT connection via a dual speed hub. Another office PC and a Linux file server were connected to the same hub. Network utilization was generally below 2%.

On average, 11 milliseconds are required to transfer a single tag, be it a scalar REAL, BOOL or DINT, an array of 15 REAL or 352 BOOL values. Since the EPICS driver combines multiple requests up to the PLC buffer limit, about 15 tags, each with a 15-character name, can be read in one transfer of about 20 ms while the separate transfers would require more than 160 ms.

In an attempt to simulate a common application, an IOC was configured with 352 binary input records, scanning the elements of a BOOL array at 10Hz, and 120 analog input records, scanning three 40-element REAL arrays at 2 Hz. Since Ethernet is not deterministic, these transfer rates vary over time due to collisions on the network, resulting in a transfer time histogram as shown in Fig. 1.

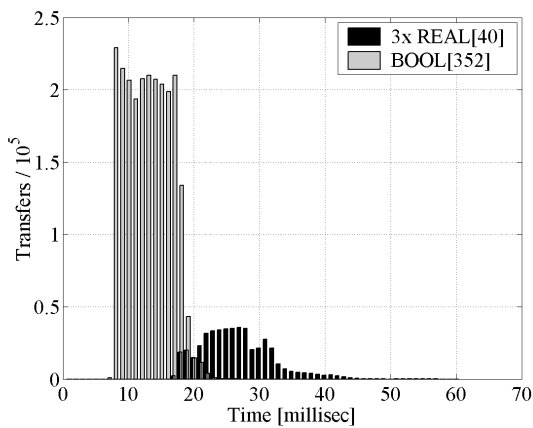

Figure 1: Transfer times sampled over 3 days.

On average, the whole BOOL array transfer was handled in 15ms, all REAL arrays were transferred in 25ms, so that the records could easily be updated at the chosen scan rate.

In these measurements, the "System Overhead Time Slice" of the PLC was set to 10%. After increasing it to 50% and connecting the CPU and PLC to a network switch, the average times for a single tag are reduced to 7ms.

## 7 CONCLUSION

It is possible to use the published EtherNet/IP specification together with the openly available Allen-Bradley extensions to read and write tags on a ControlLogix system.

The EPICS support is convenient to use. The record configuration can be changed at runtime; transfers are automatically combined up to the PLC buffer limit. In contrast to other protocols, there is no need to define the affected tags as "produced" or "consumed", nor does it require a network-wide assessment of timing values for scheduling transfers as the original ControlNet did.

Due to the nature of Ethernet, the exact transfer time varies. A switched network topology will help minimize variations. To reach the required throughput, values of interest should be arranged in arrays. To keep the ladder logic readable, intelligible tag names can be used to alias the meaningless transfer array elements.

The published CIP service codes do not allow browsing of PLC tag names and type information, which would allow an even more user-friendly EPICS driver.

With explicit messaging, each tag transfer is a round-trip request. The tested version of the 1756-ENET module does not support implicit messaging. Ideally the IOC could subscribe to the tags of interest and from then on receive asynchronous notification of changes or at least periodic updates, eliminating the need to poll.

The current implementation, however, does already allow for a successful integration of ControlLogix systems into an EPICS environment

*Work supported by the Office of Basic Energy Science, Office of Science of the US Department of Energy, and by Oak Ridge National Laboratory.*